\documentclass[10pt,prc,aps]{revtex4}   
\usepackage{graphicx}
\tighten
\begin{document}
\draft
\title{                                                     
 Spin- and isospin-polarized states of nuclear matter 
in the Dirac-Brueckner-Hartree-Fock model 
 }              
\author{            
 Francesca Sammarruca}      
\affiliation{                 
 Physics Department, University of Idaho, Moscow, ID 83844-0903, U.S.A  } 
\date{\today} 
\email{fsammarr@uidaho.edu}
\begin{abstract}
Spin-polarized isospin asymmetric nuclear matter is studied within the Dirac-Brueckner-Hartree-Fock 
approach. After a brief review of the formalism, we present and discuss the self-consistent single-particle potentials at various levels of spin and 
isospin asymmetry. We then move to predictions of the energy per particle, also under different
conditions of isospin and spin polarization. Comparison with the energy per particle in isospin symmetric
unpolarized nuclear matter shows no evidence for a phase transition to a spin ordered state, neither ferromagnetic
nor antiferromagnetic.

\end{abstract}
\pacs {21.65.+f, 21.30.Fe} 
\maketitle

\section{Introduction} 
                                                                     
Describing 
the properties of nuclear matter, especially under extreme conditions, is a topic of current interest which still presents considerable theoretical  
challenges. Of particular interest is the equation of state of matter with unequal concentrations of protons 
and neutrons, because of its many applications ranging from the physics of 
rare isotopes to the properties of neutron stars. In spite of recent and fast-growing effort, 
the density dependence of the symmetry energy is not sufficiently constrained and theoretical predictions 
show considerable model dependence.

When isospin and spin asymmetries are considered together,                                                
available constraints are even more limited and predictions regarding magnetic properties of nuclear matter 
are sometimes found to be in qualitative disagreement.
Polarization properties of neutron/nuclear matter have been studied
extensively with a variety of
theoretical methods [1-28], often with contradictory conclusions. 
In the study in Ref.~\cite{IY}, 
the possibility of phase transitions into spin ordered states
of symmetric nuclear matter was explored based on the Gogny interaction 
\cite{Gogny} and the Fermi liquid formalism. There, the appearance of 
an antiferromagnetic state (with opposite spins for neutrons and protons)
was predicted, whereas the transition to a ferromagnetic state was 
not observed. This is in contrast with predictions based on the 
Skyrme interaction \cite{I03}.

The properties of polarized neutron matter (NM), in particular, have gathered much attention lately, in conjunction with the issue of  ferromagnetic instabilities together with 
the possibility of strong magnetic fields in the interior
of rotating neutron stars.                                                          
The presence of polarization would impact neutrino cross section and luminosity, resulting into a very 
different scenario for neutron star cooling. 

There are other, equally important, motivations to undertake studies of polarized matter.
In Ref.~\cite{Sam10}, for instance, we focussed on the spin degrees of freedom of symmetric nuclear matter (SNM), having in 
mind a terrestrial scenario as a possible ``laboratory". 
We payed particular attention to the spin-dependent {\it symmetry potential}, namely the gradient between               
the single-nucleon potentials for upward and downward polarized nucleons in SNM. 
The interest around this quantity arises because of its natural interpretation as a 
spin dependent nuclear optical potential, defined in perfect formal analogy to the Lane potential \cite{Lane} for the isospin degree
of freedom in isospin-asymmetric nuclear matter (IANM). 

Whether one is interested in rapidly rotating pulsars or more conventional nuclear physics, it is important to consider 
the general case where both spin and isospin asymmetries can be present. First, neutron star matter contains a 
non-negligible proton fraction. Concerning laboratory nuclear physics, 
one way to access information related to the spin dependence of the nuclear interaction in nuclear matter
is the study of collective modes such as giant resonances. Because a spin unsaturated system is usually 
also isospin asymmetric, both degrees of freedom need to be taken into account.

In previous calculations \cite{Sam10,SK07}, we have investigated spin-polarized pure neutron matter and symmetric matter.    
The  purpose of this paper is to extend our previous predictions to include matter with different 
concentrations of neutrons and protons where each nucleon species can have definite spin polarization. 
Our framework consists of the Dirac-Brueckner-Hartree-Fock (DBHF) approach to nuclear matter together with a 
realistic meson-theoretic potential, which we choose to be the Bonn B potential \cite{Mac89}. 
To the best of our knowledge, this kind of calculation for spin polarized asymmetric nuclear matter (SPANM) 
is not in the literature.

This paper is organized as follows. In the next section we review the main aspects of the                    
procedure leading to the self-consistent determination of the one-body potentials experienced by a single nucleon in SPANM
 together with the effective interaction. 
The characteristics of those potentials are discussed in Section {\bf III}. 
We then proceed to show results for the energy/particle, namely the equation of state (EoS) of SPANM 
under extreme conditions of polarization (Section IV).
The existence (or not) of a possible phase transition can be argued by comparing the 
energies of the fully polarized and the unpolarized phases. 
A brief summary and our conclusions are contained in the last section. 

\section{Brief review of the self-consistent method } 

Our calculation is microscopic and treats nucleons
relativistically.                                                  
Within the Dirac-Brueckner-Hartree-Fock (DBHF) method, 
the interactions of the nucleons with the nuclear medium are expressed as self-energy corrections to the 
nucleon propagator. That is, the nucleons are regarded as ``dressed" quasi-particles.                                     
Relativistic effects lead to an intrinsically density-dependent interaction which is approximately consistent 
with the contribution from three-body forces (TBF) typically employed in non-relativistic approaches, particularly those TBF of the 
``Z-diagram" type, which originate from the presence of negative energy Dirac states (antinucleons).

The starting point of any microscopic calculation of nuclear
structure or reactions is a realistic free-space nucleon-nucleon interaction.      
 Our standard
framework consists of the Bonn B one-boson-exchange (OBE)potential \cite{Mac89} together with the
DBHF approach to nuclear matter. A
detailed description of our application of the DBHF method to
SNM, NM, and IANM can be found in a recent review of our 
work \cite{FS10}. (In the bibliography of Ref.~\cite{FS10} the reader will find a fairly complete list of original DBHF papers concerning SNM.) 

In a spin-polarized and isospin asymmetric system with fixed total density, $\rho$,               
the partial densities of each species are               
\begin{equation}
\rho_n=\rho_{nu}+\rho_{nd}\; , \; \; \; 
\rho_p=\rho_{pu}+\rho_{pd}\;, \; \; \; 
\rho=\rho_{n}+\rho_{p} \; ,           
\end{equation}
where $u$ and $d$ refer to up and down spin-polarizations, respectively, of protons ($p$) or neutrons ($n$). 
The isospin and spin asymmetries, $\alpha$, $\beta_n$, and $\beta_p$,  are defined in a natural way: 
\begin{equation}
\alpha=\frac{\rho_{n}-\rho_{p}}{\rho} \;, \; \; \;
\beta_n=\frac{\rho_{nu}-\rho_{nd}}{\rho_n} \;, \; \; \; 
\beta_p=\frac{\rho_{pu}-\rho_{pd}}{\rho_p} \;. 
\end{equation}
The density of each individual component can be related to the total density by 
\begin{equation}
\rho_{nu}=\frac{1 + \beta_n}{2}\frac{1 + \alpha}{2}\rho \; , 
\end{equation}
\begin{equation}
\rho_{nd}=\frac{1 - \beta_n}{2}\frac{1 + \alpha}{2}\rho \; , 
\end{equation}
\begin{equation}
\rho_{pu}=\frac{1 + \beta_p}{2}\frac{1 - \alpha}{2}\rho \; , 
\end{equation}
\begin{equation}
\rho_{pd}=\frac{1 - \beta_p}{2}\frac{1 - \alpha}{2}\rho \; , 
\end{equation}
where each partial density is related to the corresponding Fermi momentum 
through $\rho_{\tau \sigma}$ =$ \frac{(k_F^{\tau \sigma})^3}{6 \pi^2}$. 
The {\it average} Fermi momentum  and the total density are related in the usual way as 
$\rho= \frac{2 k_F^3}{3 \pi ^2}$. 

The single-particle potential of a nucleon in a particular $\tau \sigma$ state, $U_{\tau \sigma}$, is the solution of a
set of four coupled equations, 
\begin{equation}
U_{nu} = U_{nu,nu} + U_{nu,nd} + U_{nu,pu} + U_{nu,pd}       
\end{equation} 
\begin{equation} 
U_{nd} = U_{nd,nu} + U_{nd,nd} + U_{nd,pu} + U_{nd,pd}        
\end{equation} 
\begin{equation} 
U_{pu} = U_{pu,nu} + U_{pu,nd} + U_{pu,pu} + U_{pu,pd}      
\end{equation} 
\begin{equation} 
U_{pd} = U_{pd,nu} + U_{pd,nd} + U_{pd,pu} + U_{pd,pd}   \; , 
\end{equation}
to be solved self-consistently along with the two-nucleon $G$-matrix.                   
In the above equations,                                                              
each $U_{\tau \sigma, \tau '\sigma'}$ term contains the
appropriate (spin and isospin dependent) part of the interaction, $G_{\tau \sigma,
'\tau'\sigma'}$. More specifically,
\begin{equation}
U_{\tau \sigma}({\vec k}) = \sum _{\sigma '=u,d}\sum_{\tau'=n,p} \sum _{q\leq k_F^{\tau' \sigma
'}} <\tau \sigma,\tau'\sigma'|G({\vec k},{\vec q})|\tau \sigma,\tau'\sigma'>,
\end{equation}
where the third summation indicates integration over the Fermi
seas of protons and neutrons with spin-up and spin-down, and                           
\begin{widetext}
\begin{eqnarray}
<\tau \sigma,\tau'\sigma'|G({\vec k},{\vec q})|\sigma \tau,\sigma '\tau'>&=&
\sum_{L,L',S,J,M,M_L,T} |<\frac{1}{2} \sigma;\frac{1}{2} \sigma '|S
(\sigma + \sigma ')>|^2
|<\frac{1}{2} \tau;\frac{1}{2} \tau '|T
(\tau + \tau ')>|^2 
\nonumber\\
&\times&<L M_L;S(\sigma + \sigma ')|JM>
<L' M_L;S(\sigma + \sigma ')|JM> \nonumber\\
&\times& i^{L'-L} Y^{*}_{L',M_L}({\hat k_{rel}}) Y_{L,M_L}({\hat
k_{rel}}) <LSJ|G(k_{rel},K_{c.m.})|L'SJ> \; . 
\end{eqnarray}
\end{widetext}
Consistent with the DBHF method, the $G$-matrix contains medium effects from Pauli
blocking, dispersion, and modification of the spin-dependent nucleon field applied to the nucleon-nucleon 
potential. 

The need to separate the interaction by spin
components brings along angular dependence, with the result that the
single-particle potential depends also on the direction of the
momentum, although such dependence was found to be weak \cite{SK07}. The $G$-matrix equation is solved using
partial wave decomposition and the matrix elements are then summed
as in Eq.~(12) to provide the new matrix elements in the
representation needed for Eq.~(11), namely with spin and isospin components explicitely projected out. Furthermore, the
scattering equation is solved using relative and center-of-mass
coordinates, $k_{rel}$ and $K_{c.m.}$, since the former is a natural coordinate for the evaluation of the nuclear potential. Those are then easily related
to the momenta of the two particles, $k$ and $q$, in order to
perform the integration indicated in Eq.~(11).  Notice that solving
the $G$-matrix equation requires knowledge of the single-particle
potential, which in turn requires knowledge of the effective interaction.
Hence, Eqs.~(7-10) together with the $G$-matrix equation constitute a
rather lengthy self-consistency problem.                                                  
The latter starts with an {\it ansatz} for the single-particle potential as suggested by the 
most general structure of the nucleon self-energy operator consistent with all symmetry requirements. (See Ref.~\cite{FS10} and 
references therein.) 
Parametrization of the {\it ansatz} and comparison with Eq.~(11) at every step of the iterative procedure, a method known as the ``reference spectrum approximation", allow the determination of the single-nucleon
potentials in each $\tau \sigma$ channel. 

The kernel of the $G$-matrix equation contains 
the Pauli operator for scattering of two particles with
two different Fermi momenta, $k_F^{\tau \sigma}$ and 
 $k_F^{\tau' \sigma'}$, which is  
defined in analogy with the
 one for IANM \cite{AS1},
\begin{equation}
Q_{\tau \sigma, \tau' \sigma'}(k,q,k_F^{\tau \sigma},k_F^{'\tau'\sigma'})=\left\{
\begin{array}{l l}
1 & \quad \mbox{if $p>k_F^{\tau \sigma}$ and  $q>k_F^{\tau' \sigma'}$}\\
0 & \quad \mbox{otherwise.}
\end{array}
\right.
\end{equation}
The Pauli operator is then expressed in terms of $k_{rel}$ and
$K_{c.m.}$ and angle-averaged in the usual way.

Once a self-consistent solution for Eqs.~(7-11) has been obtained, the average potential energy for 
a given $\tau \sigma$ component can be calculated. 
A final average over all              
$\tau \sigma$ components provides, along with the kinetic energy $K_{\tau \sigma}$, the average energy/particle in spin-polarized
isospin-asymmetric nuclear matter. 
Specifically, 
\begin{equation}
\frac{E}{A} = \frac{1}{A} \sum _{\sigma =u,d}\sum_{\tau=n,p} \sum _{k\leq k_F^{\tau \sigma     
}} \Big (K_{\tau \sigma}(k) + \frac{1}{2} U_{\tau \sigma}(k) \Big ) \; ,         
\end{equation}
where $E/A$ is a function of $\rho$, $\alpha$, $\beta_n$, and $\beta_p$. 
We recall that, in the DBHF approach, the kinetic energy is obtained from the expectation value of the 
free-particle operator in the Dirac equation. 

All calculations are conducted including values of the total angular momentum from 0 to 6, which
we have verified to provide satisfactory convergence.

\section{One-body potentials in SPANM}                                                                  
The single-particle potential in nuclear matter 
is a very important quantity as it can be viewed as the optical potential in the interior of a nucleus and 
thus, to a certain extent, can be constrained by optical potential analyses. 
In this section, we present and discuss its dependence on the momentum and on spin/isospin asymmetries. 

\begin{figure}[!t] 
\centering 
\vspace*{-1.0cm}
\hspace*{-1.0cm}
\scalebox{0.35}{\includegraphics{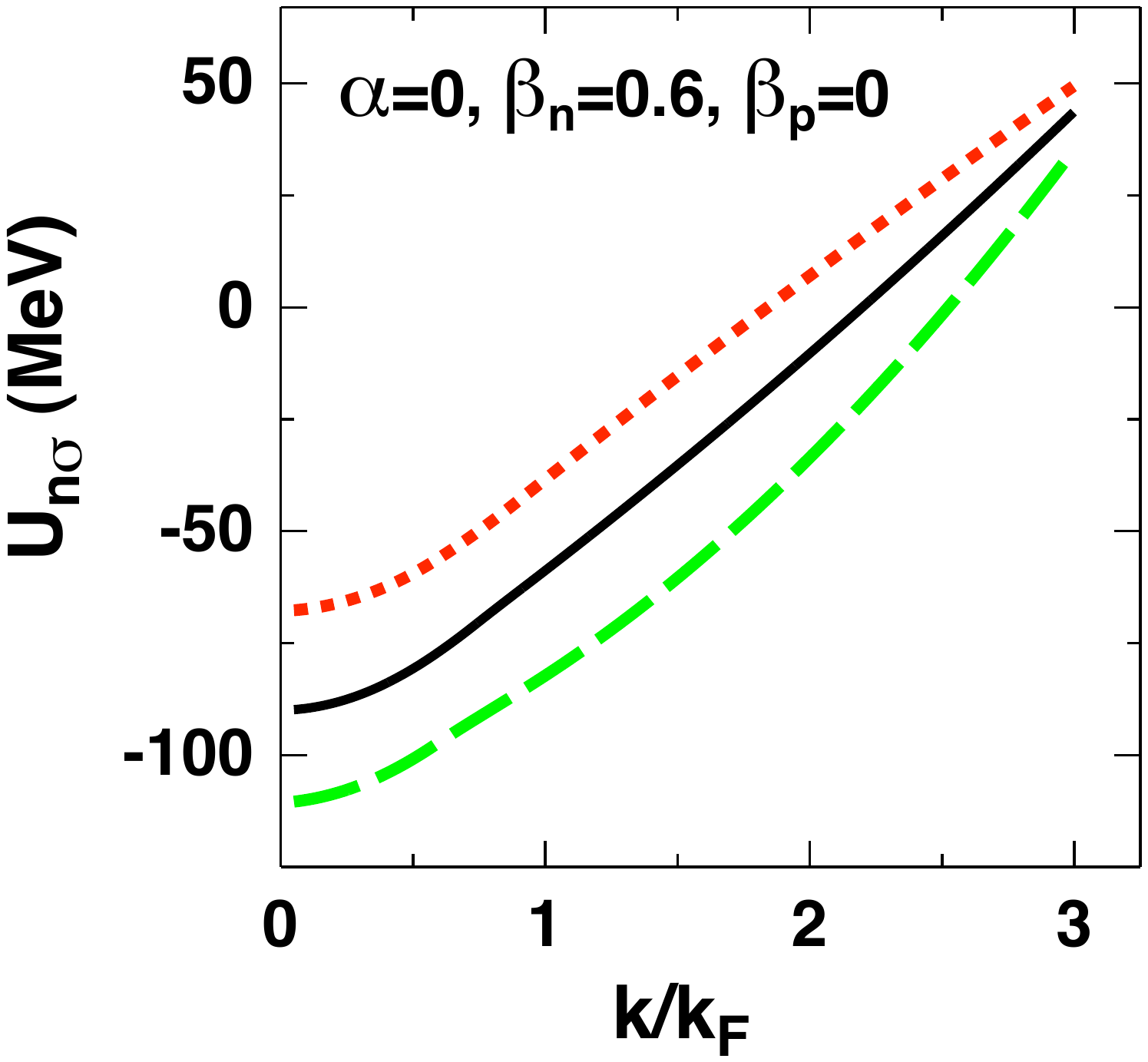}} 
\scalebox{0.35}{\includegraphics{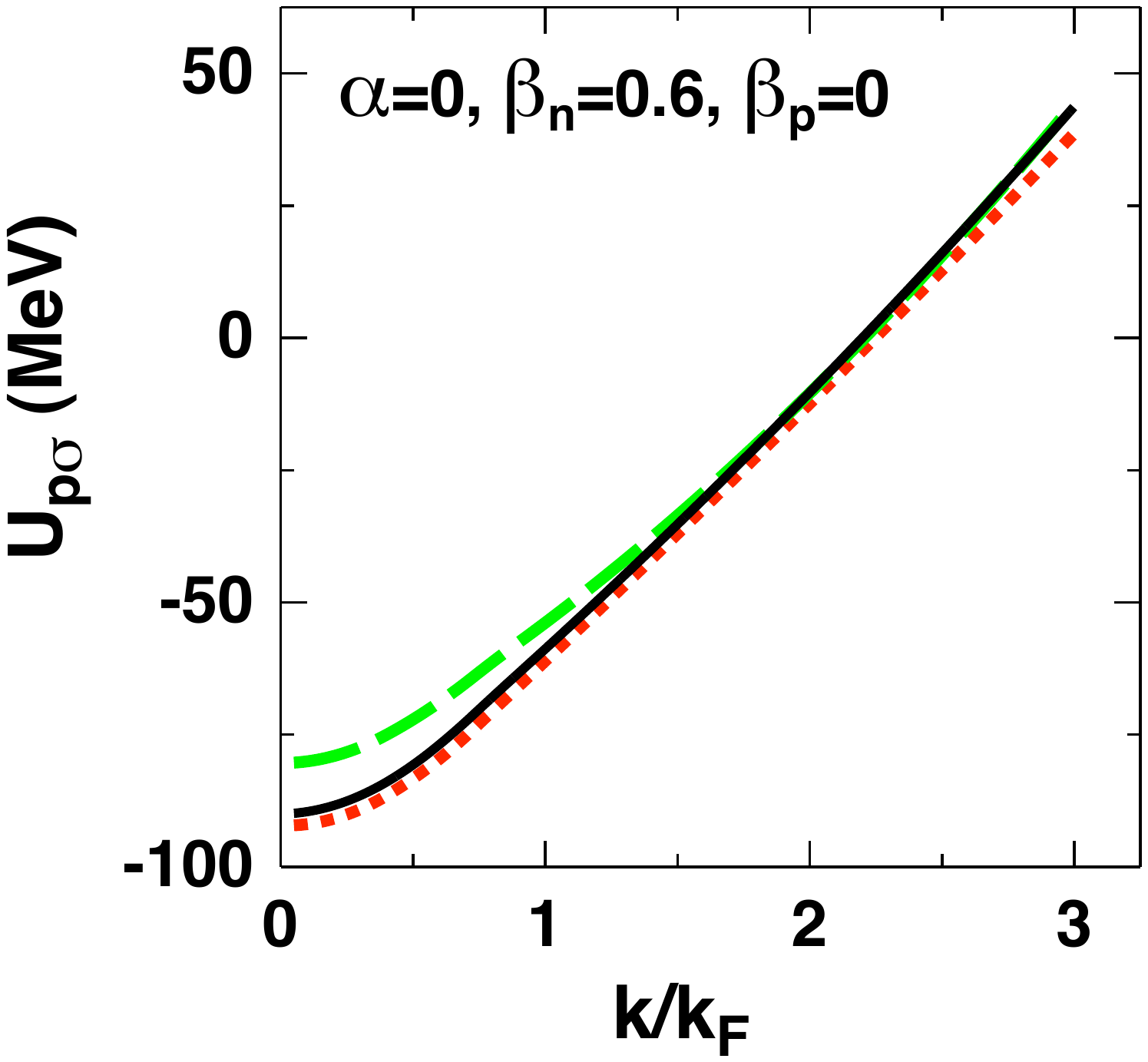}} 
\vspace*{-2.5cm}
\caption{(color online)                                        
The neutron (left frame) and the proton (right frame) single-particle potentials in isospin symmetric matter with neutron and proton polarizations
as indicated inside the frames.       
The (black) solid line is the prediction for unpolarized matter. In both frames, the  (red) dotted and (green) dashed lines   
are the predicted $U_{\tau u}$ and $U_{\tau d}$, respectively.                        
The horizontal axis is the momentum in units of the average Fermi momentum, which is equal to 1.4 fm$^{-1}$. 
} 
\label{one}
\end{figure}

\begin{figure}[!t] 
\centering 
\vspace*{-1.0cm}
\hspace*{-1.0cm}
\scalebox{0.35}{\includegraphics{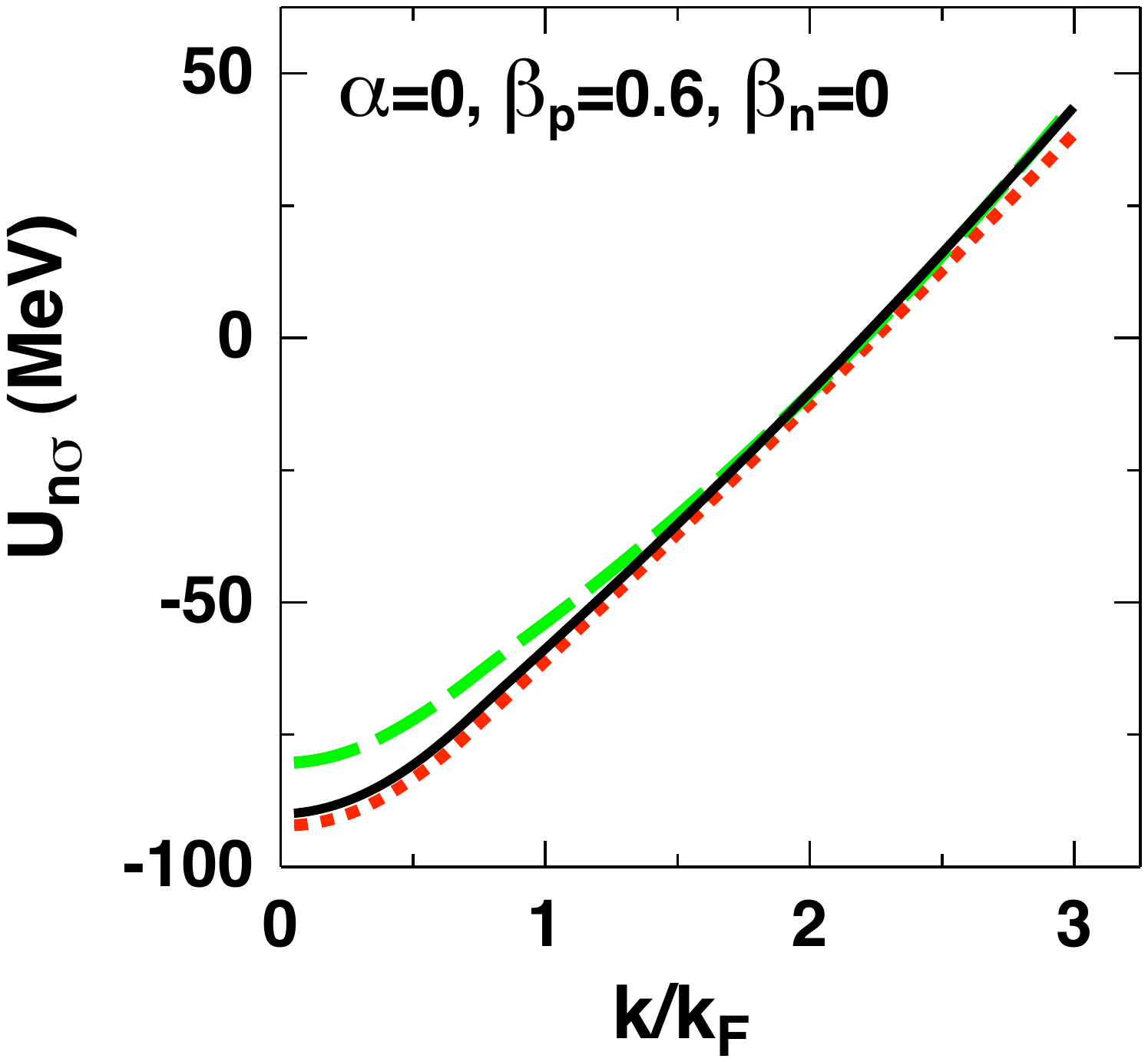}} 
\scalebox{0.35}{\includegraphics{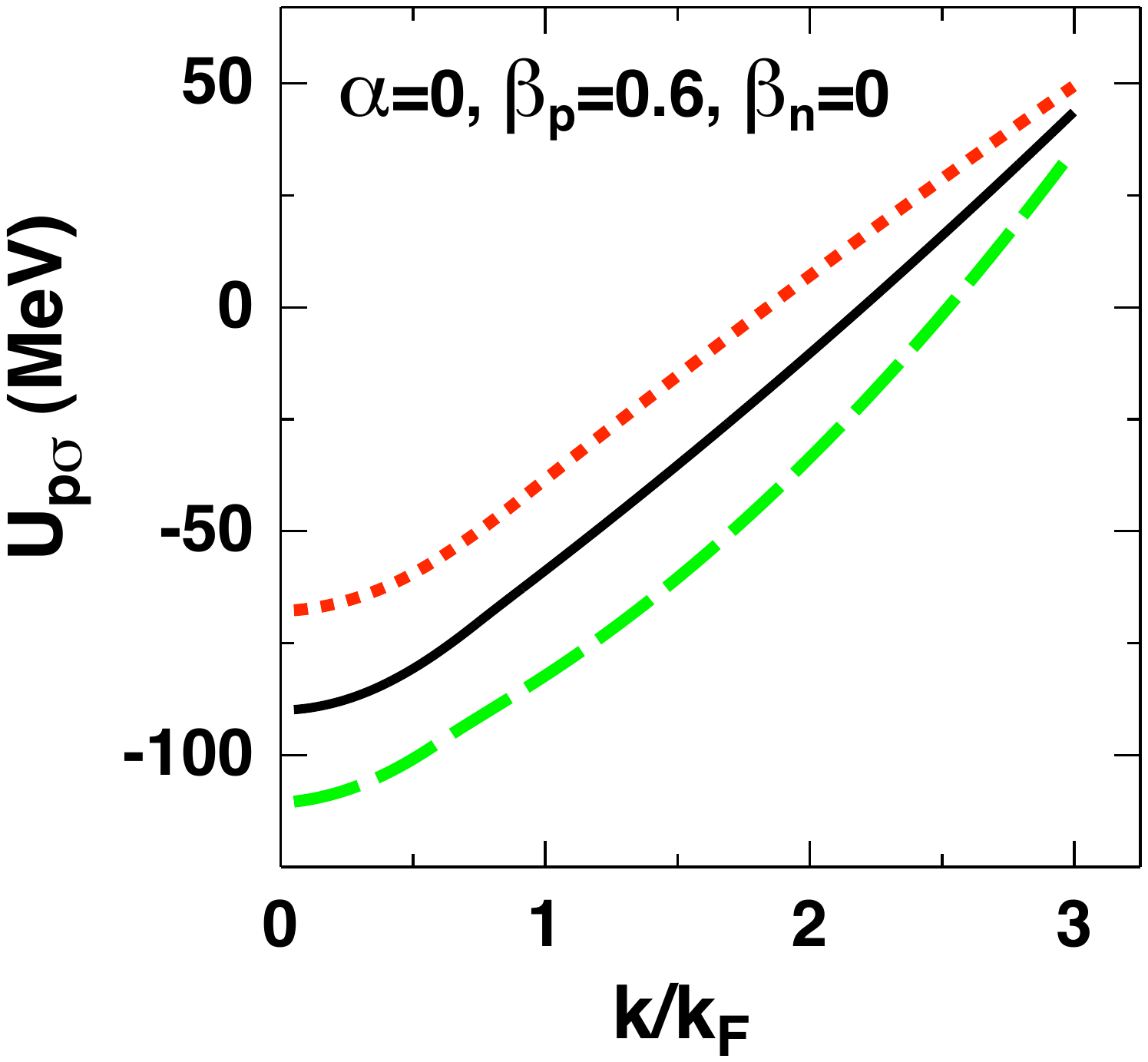}} 
\vspace*{-2.5cm}
\caption{(color online)                                        
The neutron (left frame) and the proton (right frame) single-particle potentials in isospin symmetric matter with neutron and proton polarizations
as indicated inside the figures.  
Legend as in Fig.~1. 
} 
\label{two}
\end{figure}

\begin{figure}[!t] 
\centering 
\vspace*{-1.0cm}
\hspace*{-1.0cm}
\scalebox{0.35}{\includegraphics{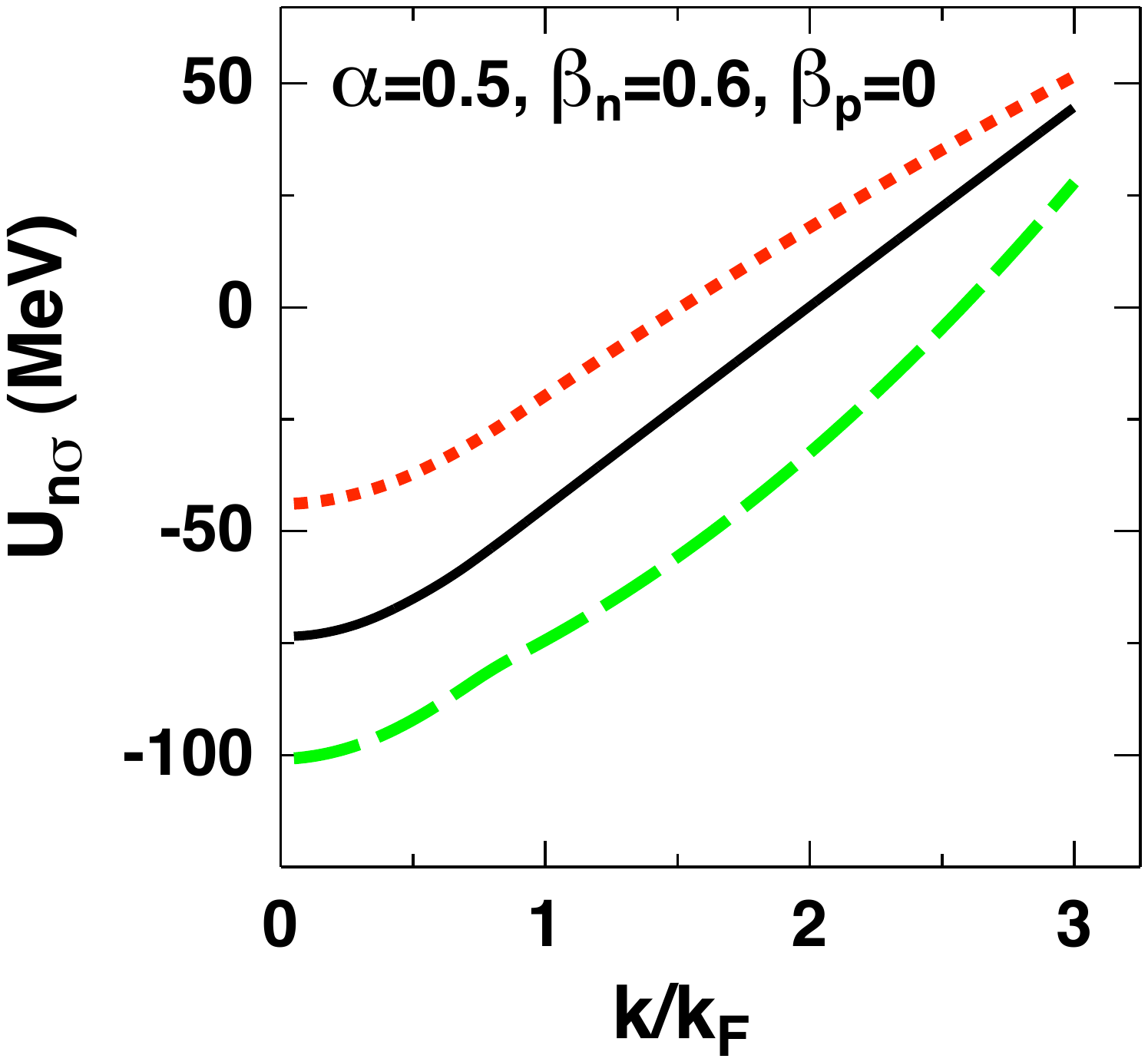}} 
\scalebox{0.35}{\includegraphics{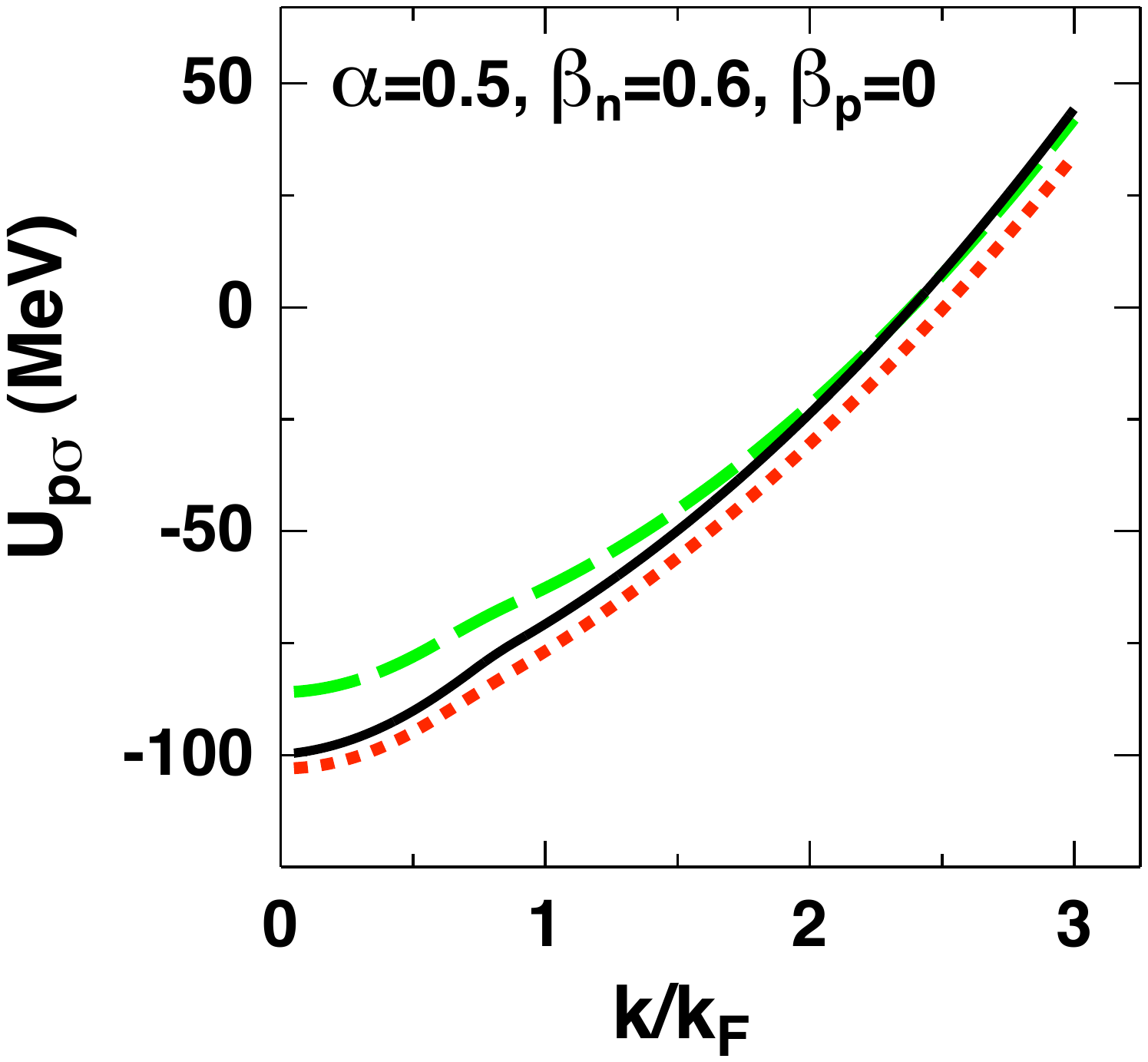}} 
\vspace*{-2.5cm}
\caption{(color online)                                        
The neutron (left frame) and the proton (right frame) single-particle potentials in isospin {\it asymmetric} matter with neutron and proton polarizations
as indicated inside the figures.  
Legend as in Fig.~1. 
} 
\label{three}
\end{figure}

\subsection{Momentum dependence}                                                                  
In Fig.~1 (left panel), we show the momentum dependence of the one-body potentials for upward and downward polarized 
neutrons in isospin symmetric ($\alpha$=0) nuclear matter. The protons are unpolarized whereas the neutron 
spin polarization parameter is taken to be 0.6. The right panel of Fig.~1 shows the same quantity for protons. 
In both cases, the solid curve is the prediction of the single-particle potential in unpolarized matter. 
All potentials are calculated at a density equal to 0.185 fm$^{-3}$. In all cases, the polar angle of the momentum 
vector ${\vec k}$ is taken to be zero. 

With a larger number of upward-polarized neutrons, $U_{nu}(k)$ becomes more repulsive while 
 $U_{nd}(k)$ turns more attractive. Notice that the opposite trend is displayed by 
 $U_{pu}(k)$ and                                                                       
 $U_{pd}(k)$.                                                                          
The reason for the observed splittings is of course in the spin dependence of the $G$-matrix (and isospin 
dependence, when applicable), together with the fact 
that the number of interactions a single nucleon (with 
specified $\tau \sigma$) can undergo with other ($\tau' \sigma'$) nucleons 
changes as the population of one species increases or decreases.                                        

Figure 2 shows a situation parallel to the one presented in Fig.~1, except that the neutrons are now unpolarized. 
Comparison between Fig.~1 and Fig.~2 shows that, 
as it can be expected, the role of neutrons and protons are perfectly interchanged when 
$\beta_n\rightarrow \beta_p$ and 
$\beta_p\rightarrow \beta_n$.       

In Fig.~3, we investigate the impact of including isospin 
asymmetry as well, specifically a neutron excess given by $\alpha$=0.5.  
(Notice that the neutron and proton potentials in absence of polarization (solid curves in both frames of Fig.~3)   
are different to begin with due to the isospin asymmetry.) 
The splitting remains qualitatively similar to the case of Fig.~1, but it is more pronounced for the nucleon
type whose density is larger. 

In all cases, the momentum dependence remains qualitatively similar to the one displayed in 
unpolarized SNM, with the size of the splitting larger at the lower momenta, which may be                
due to weaker sensitivity of a high-momentum nucleon to medium and asymmetry effects. 

\begin{figure}[!t] 
\centering 
\vspace*{-1.0cm}
\hspace*{-1.0cm}
\scalebox{0.35}{\includegraphics{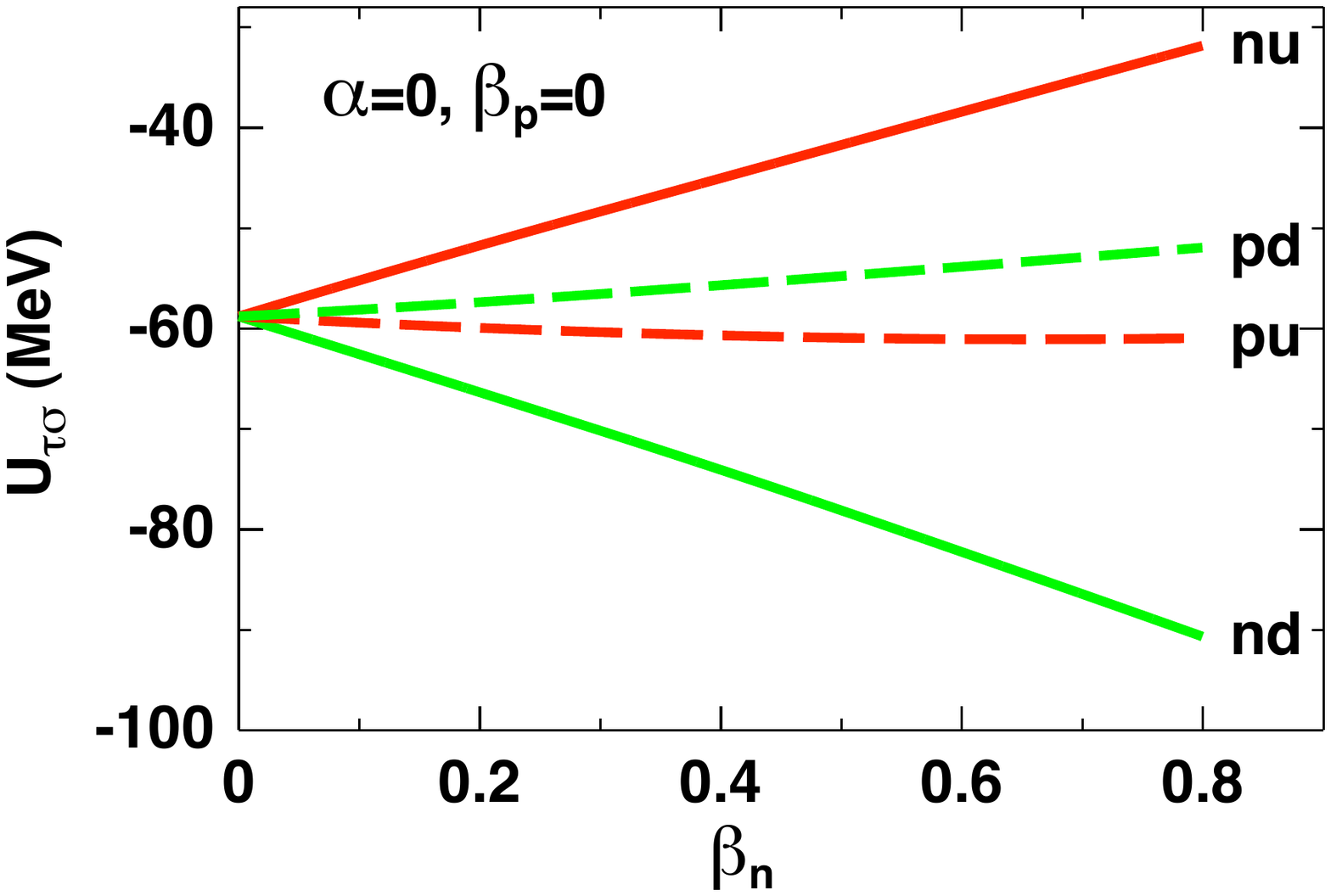}} 
\scalebox{0.35}{\includegraphics{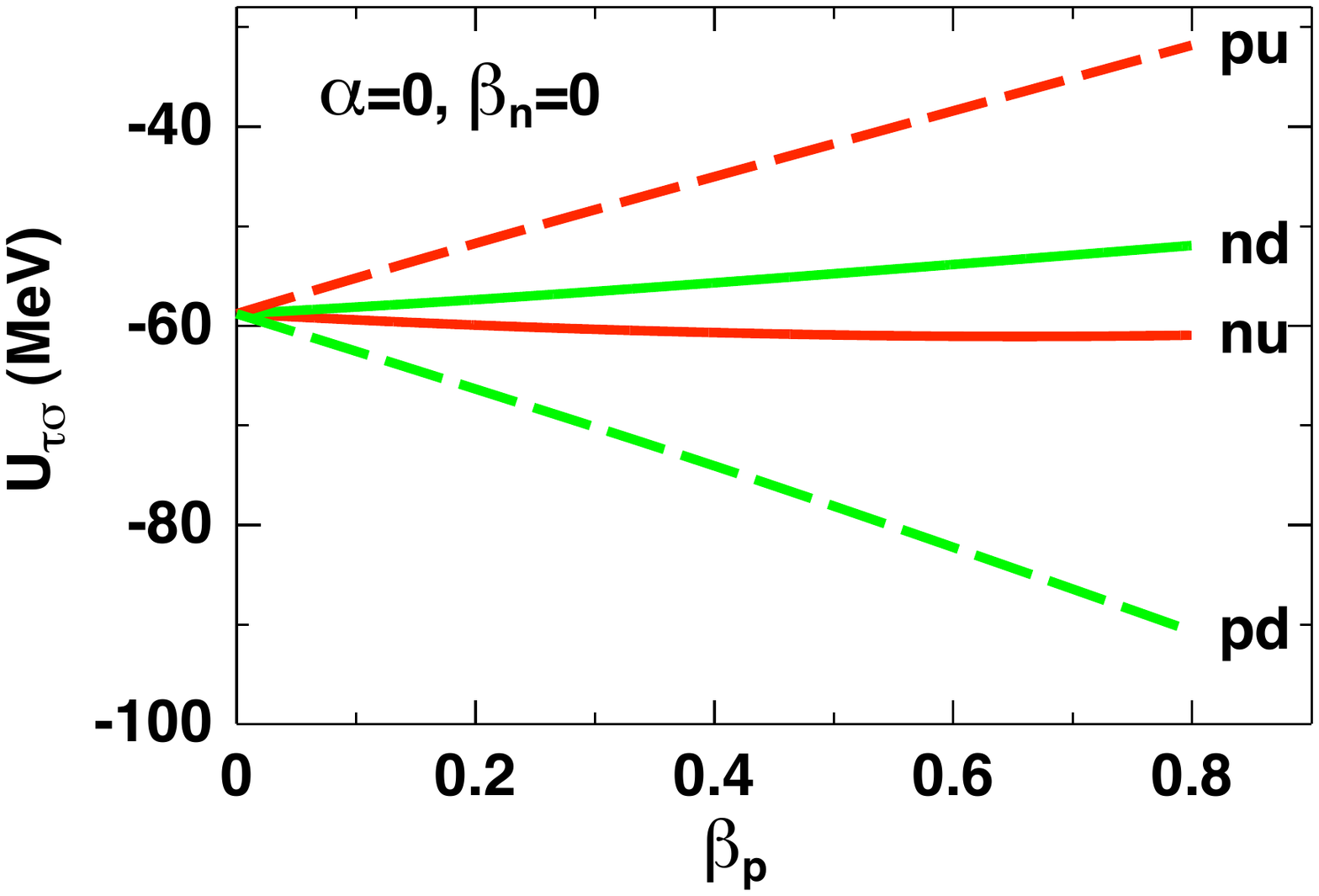}} 
\vspace*{-2.5cm}
\caption{(color online)                                        
The neutron (solid line) and the proton (dashed lines) single-particle potentials in isospin symmetric matter {\it vs.} the 
neutron (left panel) and the proton (right panel) spin polarizations. The nucleon momentum is fixed and equal to 
the average Fermi momentum, 1.4 fm$^{-1}$. 
} 
\label{four}
\end{figure}
\begin{figure}[!t] 
\centering 
\vspace*{-1.0cm}
\hspace*{-1.0cm}
\scalebox{0.35}{\includegraphics{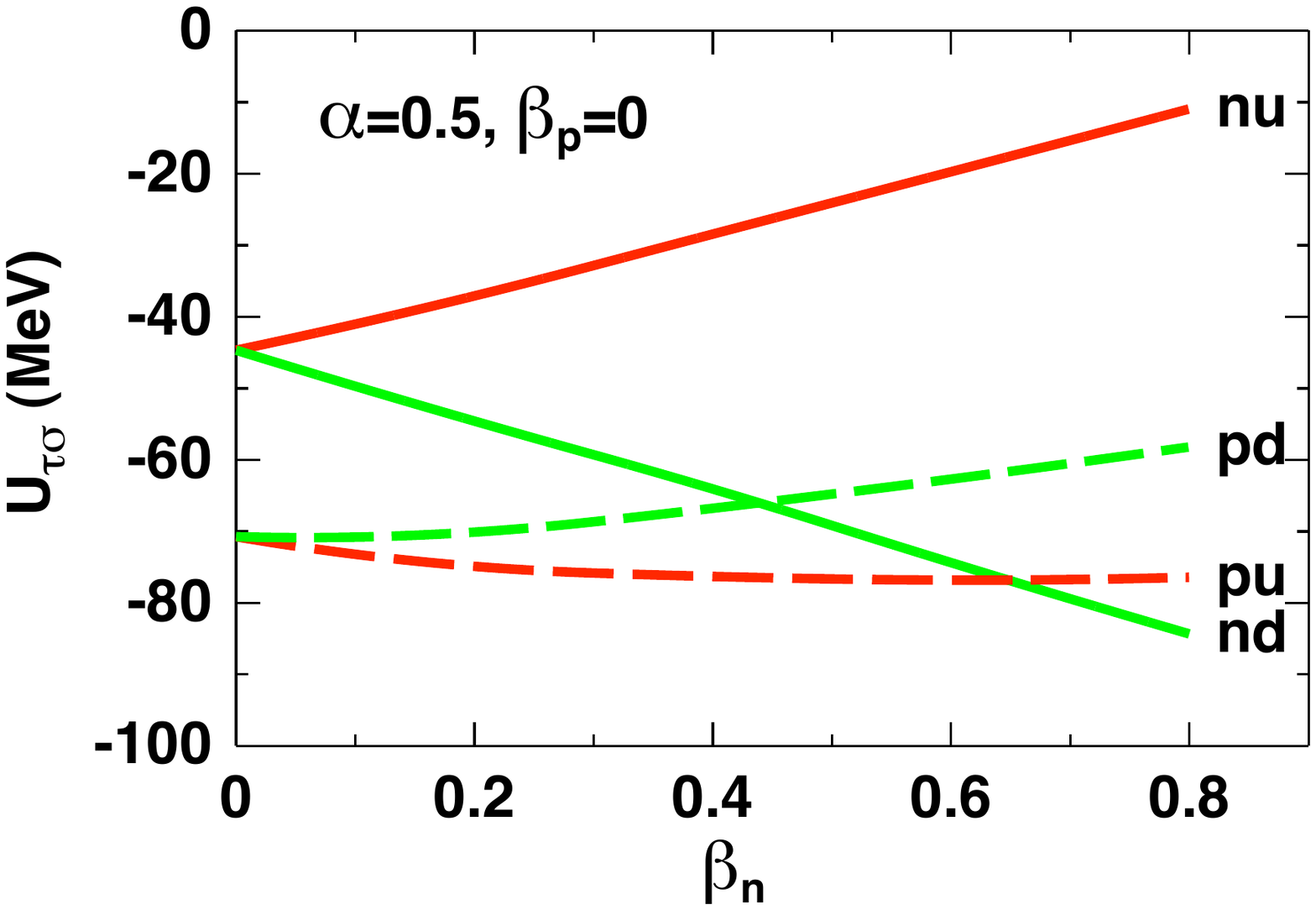}} 
\scalebox{0.35}{\includegraphics{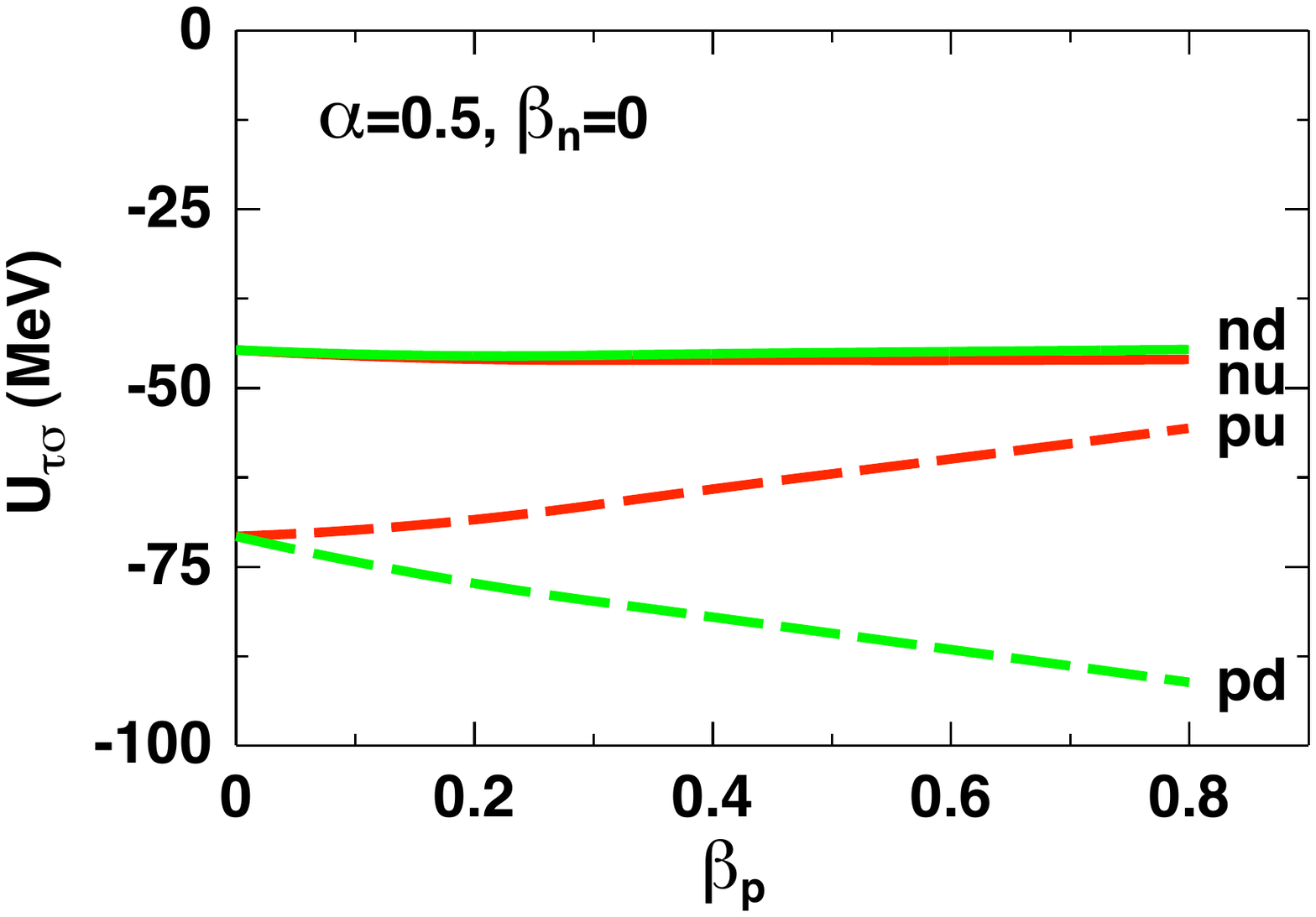}} 
\vspace*{-2.5cm}
\caption{(color online)                                        
The neutron (solid) and the proton (dashed) single-particle potentials in isospin asymmetric matter {\it vs.} the 
neutron (left panel) and the proton (right panel) spin polarizations. The nucleon momentum is fixed and equal to 
the average Fermi momentum, 1.4 fm$^{-1}$. 
} 
\label{five}
\end{figure}

\subsection{Spin and isospin asymmetry dependence}                                                                  

Here we focus on the dependence of the single-particle potentials on various levels of asymmetries, for
fixed total density and momentum ${\vec k}$. 
First, we show the splitting of the single-neutron and single-proton potentials in isospin symmetric 
matter with changing neutron polarization (and for zero proton polarization), see left frame of Fig.~4. 
The right frame of Fig.~4 confirms that the appropriate symmetry is respected when neutron and proton polarizations are
interchanged. 

For the predictions of Fig.~5, isospin asymmetry has been introduced as well. Notice that the predictions shown in Fig.~5 
are not, and should not be symmetric with respect to $n\leftrightarrow p$ exchange. This is the case
if $\alpha \rightarrow -\alpha$, in addition to   
 $\beta_n \rightarrow \beta_p$ and                 
 $\beta_p \rightarrow \beta_n$ (that is, under charge exchange). 

As pointed out in the previous subsection, the size and direction of the various splittings depend sensitively on the 
strength of the partial contributions, $U_{\tau \sigma, \tau' \sigma '}$, to each 
 $U_{\tau \sigma}$ potential, see Eqs.~(7-11), 
which in turn receive contributions from $G$-matrix elements in different spin and isospin channels. 
We will come back to this point in the next section when discussing the energy/particle. 

The curves displayed in Figs.~4-5 show an approximately linear behavior, although some 
deviations from linearity can be seen, especially for the weaker potentials in asymmetric matter. We observed a similar
trend when only isospin splitting or only spin splitting (in NM or SNM) 
was considered \cite{Sam10}. 

\section{Energy per particle in SPANM}                                                                  
When the potential and kinetic energies are averaged as in Eq.~(14), one obtains
the energy/nucleon for a given state of isospin asymmetry and spin polarization. 
To render the four-dimensional self-consistent calculation more manageable, we ignore the angular dependence, which
was found to be weak both in nuclear and in neutron matter \cite{pol16,SK07}, and keep the polar angle of the nucleon momentum vector
at a constant value, for which we choose            
$\pi/4$. We have tested this choice in a few cases and found it to give good agreement with the result of                
averaging over all angles. 
(Notice that 
 single-nucleon potentials in polarized matter have their maximum or minimum values at either zero or 
$\pi/2$.)                                 

In the left panel of Fig.~6, we show, 
in comparison with unpolarized symmetric matter (solid line):            
the EoS 
for the case of fully polarized neutrons and completely unpolarized protons (dashed line); 
the EoS 
for the case of protons and neutrons totally polarized in the same direction, that is, matter in the 
ferromagnetic (FM) state ( dashed-dotted line); 
the EoS 
for the case of protons and neutrons totally polarized in opposite directions, namely matter in the antiferromagnetic 
(AFM) state ( dotted line). 
A similar comparison is shown in the right panel of Fig.~6, but for isospin asymmetric matter. 
(Notice that all predictions  
are invariant under a global spin flip, as we have verified directly.) 

\begin{figure}[!t] 
\centering 
\vspace*{-1.0cm}
\hspace*{-1.0cm}
\scalebox{0.35}{\includegraphics{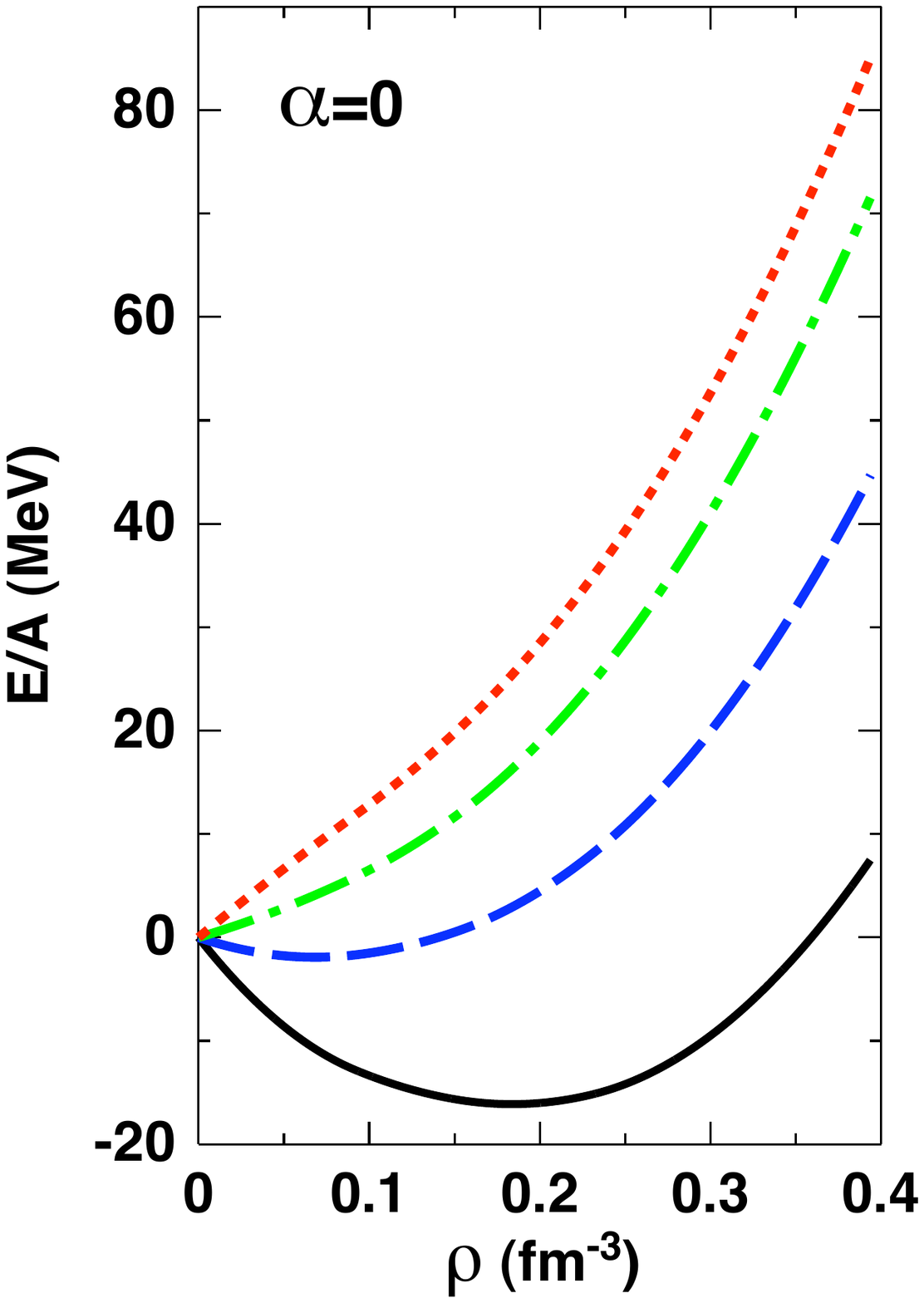}} 
\scalebox{0.35}{\includegraphics{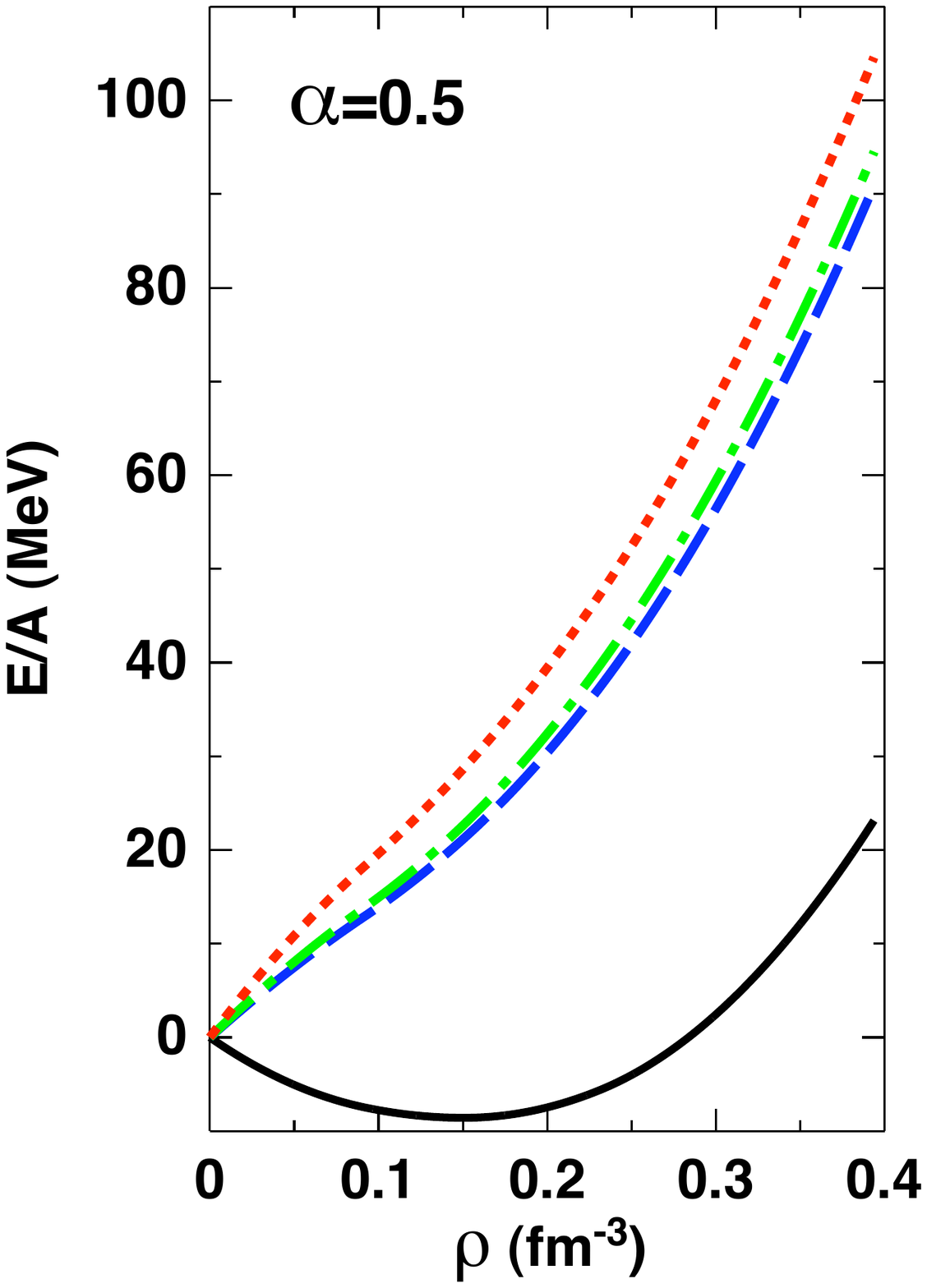}} 
\vspace*{-1.5cm}
\caption{(color online)                                        
The energy per particle as a function of density and variuos degrees of proton and neutron polarizations
in symmetric matter (left) and asymmetric matter (right). In both frames, the (blue) dashed line corresponds to 
totally polarized neutrons and unpolarized protons ($\beta_n$=1, $\beta_p$=0); the (green) dash-dotted line is the 
prediction for the FM state 
($\beta_n$=1, $\beta_p$=1); the (red) dotted line shows the energy of the AFM state 
($\beta_n$=1, $\beta_p$=-1).                                                         
The (black) solid line shows the predictions for unpolarized matter. 
} 
\label{six}
\end{figure}

To better understand our findings,           
we have examined the contributions to the potential energy from singlet and triplet states separately. 
Taking as example the case of unpolarized asymmetric matter, 
up to the densities considered here (about 0.4 fm$^{-3}$) the contribution to the potential energy from singlet states was found 
to be attractive. Such contribution is absent in the fully polarized case, implying increased repulsion in the 
latter. 
Concerning triplet states, we found their contribution to the potential energy to be more repulsive in the fully
polarized case as compared to the unpolarized one. 
A similar analysis can explain the origin of the larger energy in the AFM state as compared to the unpolarized one. 

In summary, we find that, for both symmetric and asymmetric matter,                 
the energies of the FM and AFM states are higher than those of the corresponding unpolarized cases, with the 
AFM state being the most energetic. 
Thus, a phase transition is not anticipated in our model.               
This conclusion seems to be shared by predictions of microscopic models, such as those based on conventional Brueckner-Hartree-Fock theory 
\cite{pol18}. On the other hand, calculations based on various parametrizations of Skyrme forces result in 
different conclusions. For instance, 
 with the {\it SLy4} and {\it SLy5} forces and the Fermi liquid 
formalism                                                                                          
  a phase transition  to the AFM state is predicted in asymmetric matter 
at a critical density equal to about 2-3 times normal density \cite{IY}.

It is interesting to observe that models based on 
realistic nucleon-nucleon potentials, whether relativistic or non-relativistic, are at least in qualitative 
agreement with one another in predicting 
more energy for totally polarized states (FM or AFM) 
 up to densities well above normal density. 

On the other hand, 
qualitative disagreement is encountered with non-microscopic approaches \cite{IY} and also with 
relativistic Hartree-Fock models based on effective nucleon-meson Lagrangians. For instance, in Ref.~\cite{pol12} 
it was reported that the onset of 
a ferromagnetic transition in neutron matter, and its critical density, are crucially determined by the inclusion of isovector mesons and the 
nature of their couplings. 
Notice that our microscopic model also includes the isovector mesons $\pi$, $\rho$, and $\delta$ ($a_0$), but does 
not predict a similar scenario. The reason for this difference is most likely due to the fact that in our model all
meson-nucleon couplings are constrained by a fit to the free-space nucleon-nucleon data. In relativistic Hartree-Fock models no
such constraints are applied.

\section{Conclusions}                                                                  

Continuing with our broad analysis of nuclear matter and its extreme states, we have extended our
framework and gone beyond existing predictions.                                  
As usual, we adopt the microscopic approach for our nuclear matter calculations. 
Concerning our many-body method, we find              
DBHF to be a good starting point to look beyond the normal states of nuclear matter, which  it describes
successfully. The main strength of this method is its inherent ability to effectively incorporate 
crucial TBF contributions through relativistic effects.                                            

In this paper, we extended previous calculations to incorporate the general case of spin and isospin unsaturated matter.  
Our main result is that 
we do not predict, or forsee,       
a phase transition to a ferromagnetic or antiferromagnetic state.                                       
 In microscopic models one starts with the bare interaction and includes                      
correlations through the $G$-matrix calculation, where all important meson contributions are                               
constrained by free-space data.                          
The handling of spin and isospin dependent amplitudes, in particular whether they are tightly constrained or not, 
is most likely at the origin of the 
 divergence of predictions between microscopic and non-microscopic approaches.                                                                                             

In the near future, 
we hope to construct a                    
 convenient and sufficiently accurate parametrization of our $\rho$, 
$\alpha$, $\beta_n$, and $\beta_p$ dependent EoS. 
This may be helpful for application purposes, given that the self-consistency 
problem can be time consuming.

We point out that empirical constraints are desirable to test predictions of the                      
spin and isospin dependence of nuclear matter properties. At normal densities, systematic                                          
analyses of spin and isospin dependent optical potentials can help constraint $U_{\sigma \tau}$.

\section*{Acknowledgments}
Support from the U.S. Department of Energy under Grant No. DE-FG02-03ER41270 is 
acknowledged.                                                                           

\end{document}